\documentclass{article}

\usepackage{color}
\usepackage{amsmath, amssymb}

\newcommand{\RM}{\mathbb{R}}
\newcommand{\ZM}{\mathbb{Z}}

\newtheorem{prop}{Proposition} 
\newtheorem{cor}{Corollary}

\begin{document}

\title{{\bf Linear extrapolation for the graph of function of single variable based on walks}
\vspace{15mm}}

\author{Norio KONNO \\
Department of Applied Mathematics, Faculty of Engineering \\ 
Yokohama National University \\
Hodogaya, Yokohama, 240-8501, Japan \\
e-mail: konno-norio-bt@ynu.ac.jp \\
%Tel.: +81-45-339-4205, Fax: +81-45-339-4205 \\  \\ 
\\ 
Shohei KOYAMA \\
Graduate School of Science and Engineering \\ 
Yokohama National University \\
Hodogaya, Yokohama, 240-8501, Japan \\
e-mail: koyama-shohei-mw@ynu.jp \\
}

\date{\empty }

\maketitle

\vspace{50mm}

%\clearpage

\vspace{20mm}

%\par\noindent
%{\bf Keywords}: zeta function, quantum walk, Grover walk, regular graph, 
%integer lattice, Konno-Sato theorem 
%\par\noindent
%{\bf Concise title}: Grover/Zeta Correspondence

%\begin{small}
%\par\noindent
%{\bf Corresponding author}: Norio Konno, Department of Applied Mathematics, 
%Faculty of Engineering, Yokohama National University, Hodogaya, Yokohama, 
%240-8501, JAPAN, \\
%e-mail: konno-norio-bt@ynu.ac.jp, Tel.: +81-45-339-4205, Fax: +81-45-339-4205
%\par\noindent
%{\bf Abbr. title:} Grover/Zeta Correspondence
%\par\noindent
%{\bf Mathematical Subject Classification (2000)}: 60F05, 05C50, 15A15, 05C25. 
%�i�g�����Ƃ��͂������g���j
%\par\noindent
%{\bf PACS}: 03.67.Lx, 05.40.Fb, 02.50.Cw
%\end{small}

%\par\noindent
%{\bf 2000 Mathematical Subject Classification}: 60F05, 05C50, 15A15, 05C25. 

%\vspace{5mm}

%The contact author for correspondence: 

%Iwao Sato 

%Oyama National College of Technology, 
%Oyama, Tochigi, 323-0806, JAPAN

%Tel: +81-285-20-2176

%Fax: +81-285-20-2880

%E-mail: isato@oyama-ct.ac.jp 

\clearpage

\begin{abstract}
The quantum walk was introduced as a quantum counterpart of the random walk and has been intensively studied since around 2000. Its applications include topological insulators, radioactive waste reduction, and quantum search. The first author in 2019 defined a time-series model based on the measure of the ``discrete-time" and ``discrete-space" quantum walk in one dimension. Inspired by his model, this paper proposes a new model for the graph of a function of a single variable determined by the measure which comes from the weak limit measure of a ``continuous-time or discrete-time" and ``discrete-space" walk. The measure corresponds to a ``continuous-time" and ``continuous-space" walk in one dimension. Moreover, we also presents a method of a linear extrapolation for the graph by our model.
\end{abstract}

%----------------------------------------------------------------------

% 1.
%%%%%%%%%%%%%%%%%%%%%%%%%%%%%%%%%%%%%%%%%%%%%%%%%%%%%%%%%%%%%%%%%%%%%%%%%%%%%%
\section{Introduction \label{sec1}}
The quantum walk (QW) is a mathematical model that has been intensively investigated since around 2000 as a quantum version of the random walk (RW). The related fields include topological insulators, radioactive waste reduction, and quantum search.
The well-known properties of the QW, which are not found in the RW, are ``linear spread" and ``localization''. The linear spread means that the standard deviation of location of the QW at time $t$ is proportional to $t$, whereas that of the RW at time $t$ is proportional to $\sqrt{t}$. Localization is the phenomenon that a QW remains at the starting point even after long-time limit (see	 \cite{MW, Por, Ven}, for example).

%Let $\RM$ be the set of real numbers, $\ZM$ be the set of integers, $\RM_\geq = [0, \infty), \ \RM_> = (0, \infty),\ \ZM_\geq = \{ 0, 1, 2, \ldots \}$, and $\ZM_{>}=\{1,2,3,\ldots\}$. 
The first author in 2019 \cite{TSA} introduced a time-series model based on the discrete-time QW (DTQW) on $\ZM$, where $\ZM$ is the set of integers. His model is determined by the probability measure of the ``discrete-time" and ``discrete-space" QW. In this paper, inspired by the model, we introduce a new model for the graph of a function of a single variable, $y=f(x)$, determined by the measure which comes from the weak limit measure of a ``continuous-time or discrete-time" and ``discrete-space" QW and RW.  The measure corresponds to ``continuous-time'' and ``continuous-space'' QW and RW. More precisely, we present an evaluation function and consider the estimated value at $b(>\!\!a)$ of a graph of $y=f(x)$ on the interval $[0,a]$ using the measure obtained by the weak limit theorem for the ``continuous-time or discrete-time" and ``discrete-space" walk in one dimension. We call such an expectation ``extrapolation'' here. A typical example is to extrapolate linearly the value at $b(>\!a)$ by the Brownian motion on $\RM$ derived from the discrete-time RW (DTRW) on $\ZM$, where $\RM$ is the set of real numbers.

The rest of the present paper is as follows. Section 2 gives a definition of our model and some examples of the walk such as continuous-time QW (CTQW), DTQW, and RW. Section 3 is devoted to the evaluation function for each walk. In Section 4, we consider evaluation functions for special cases. Section 5 shows evaluation functions for some graphs of one-variable functions. Section 6 deals with the relation between the previous model \cite{TSA} and our model. Finally, in Section 7, we compute the expectation of $y=f(x)$ at $x=b$ by the parameter minimizing the evaluation function.

% 2.
%%%%%%%%%%%%%%%%%%%%%%%%%%%%%%%%%%%%%%%%%%%%%%%%%%%%%%%%%%%%%%%%%%%%%%%%%%%%%%%
\section{Definition\label{sec2}}
\indent For given and fixed $x\in\RM_\geq$ and $p\in[0,1]$, we introduce a probability measure on $\RM$, $\left\{ \mu_x(y,p) : y\in\RM \right\}$, where $\RM_\geq = [0, \infty)$. In addition, we assume that 
\begin{align}\label{mean}
\int_\RM y\ \mu_x(y,p) dy = (1-2p)x,
\end{align}
for $x\in\RM_\geq$ and $p\in[0,1]$. As for the variance of $\mu_x(y,p)$ for the walk we consider, see Eqs. (\ref{variance ctqw}), (\ref{variance dtqw}), and (\ref{variance rw}). This measure $\mu_x(\cdot,p)$ on $\RM$ can be obtained by the weak limit measure $\mu(\cdot)$ on $\RM$ for a ``continuous-time or discrete-time"  walk $X_t$ on $\ZM$ at time $t$ such as
\begin{align*}
\frac{X_t}{t^\alpha}\ \Rightarrow\ \mu  \qquad (t \to \infty),
\end{align*}
for a suitable scaling parameter $\alpha>0$. Here, $\Rightarrow$ means the weak convergence. In some case, the parameter ``location $x$'' can be interpreted as ``time $t$". From now on, we present explicit expressions $\{\mu_x(\cdot,p)\}$ of three walks, i.e., CTQW, DTQW, and RW considered in this manuscript. Remark that the range of $p$, i.e., $[0,1]$, will be extended to $\RM$ depending on situations.\\
%Put a function $f:\RM_\geq \to \RM$. Next, for a fixed $a\in \RM_>$, we define the one-dimensional graph on $[0,a]$ by $T_a = \{ y=f(x) : x\in[0,a] \}$. Then, we assume that $T_a$ is determined by a set of measures on $\RM$, $\left\{ \mu_x(y,p) : x\in \RM_\geq,\ y\in\RM,\ p\in[0,1] \right\}$, satisfying
%\begin{align}\label{mean}
%\int_\RM y \mu_x(y,p) dy = (1-2p)x,
%\end{align}
%which is given by the weak limit measure $\{\mu_x(y,p) :  x\in \RM_\geq,\ y\in\RM,\ p\in[0,1] \}$ of a walk $\{X_x : x\in\RM_\geq \}$ such as 
%\begin{align*}
%\frac{X_x}{x^\alpha}\ \Rightarrow\ \mu  \qquad (x \to \infty),
%\end{align*}
%for a suitable scaling parameter $\alpha>0$.

\textbf{Example 1}. CTQW\\
For this model, Konno \cite{limit theorem CTQW} proved the following weak limit theorem:
\begin{align}\label{limit theorem CTQW}
\frac{X_t^{(CTQW)}}{t}\ \Rightarrow\ \frac{1}{\pi\sqrt{1-y^2}}I_{(-1,1)}(y)dy \qquad (t \to \infty),
\end{align}
where the indicator function $I_A$ of $A \subset \RM$ is defined by
\begin{equation*}
I_{A}(y) = 
\begin{cases}
1   &   (y\in A),  \\
0   &   (y\notin A).
\end{cases}
\end{equation*}
Combining Eq. (\ref{mean}) with Eq. (\ref{limit theorem CTQW}), we give $\mu_x(y,p)$ as follows:
\begin{align*}
\mu_x(y,p) = \frac{1}{\pi x \sqrt{1-\left(\frac{y-c}{x}\right)^2}}I_{(c-x,c+x)}(y) \qquad (x\in \RM_>,\ y\in\RM,\ p\in[0,1]),
\end{align*}
where $c=(1-2p)x$ and $\RM_> = (0, \infty)$.\\

\textbf{Example 2}. DTQW\\
As in the case of Example 1, we use the following weak limit theorem shown by Konno \cite{limit theorem DTQW 2002, limit theorem DTQW 2005}:
\begin{align}\label{limit theorem DTQW}
\frac{X_t^{(DTQW)}}{t}\ \Rightarrow\ \frac{\sqrt{1-r^2}}{\pi  (1-y^2)\sqrt{r^2-y^2}}I_{(-r,r)}(y)dy \qquad (t \to \infty),
\end{align}
where $r \in (0,1)$.
We should note that if $r=1/\sqrt{2}$, then the DTQW becomes the Hadamard walk which is one of the most investigated models in the study of the QW. From Eqs. (\ref{mean}) and  (\ref{limit theorem DTQW}), we present $\mu_x(y,p)$ in the following:
\begin{align*}
\mu_x(y,p) = &\frac{\sqrt{1-r^2}}{\pi x \left(1-\left(\frac{y-c}{x}\right)^2\right)\sqrt{r^2-\left(\frac{y-c}{x}\right)^2}}I_{(c-xr,c+xr)}(y) \\
& \hspace{60mm} (x\in \RM_>,\ y\in\RM,\ p\in[0,1]),
\end{align*}
where $c=(1-2p)x$.\\

\textbf{Example 3}. RW (= continuous-time RW (CTRW), DTRW)\\
As for the CTRW and DTRW, the following result on the weak limit theorem is well-known (see Proposition 1.2.1 in \cite{Lawler and Limic}, for example):
\begin{align}\label{limit theorem RW}
\frac{X_t^{(CTRW)}}{\sqrt{t}},\ \frac{X_t^{(DTRW)}}{\sqrt{t}} \ \Rightarrow\  \frac{1}{\sqrt{2\pi}}e^{-\frac{y^2}{2}}dy \qquad (t \to \infty),
\end{align}
for $y\in\RM$.
It follows from Eqs. (\ref{mean}) and (\ref{limit theorem RW}) that we have
\begin{align*}
\mu_x(y,p) =\frac{1}{\sqrt{2\pi x}}e^{-\frac{\left( y-c \right)^2}{2x}} \qquad (x\in \RM_>,\ y\in\RM,\ p\in[0,1]),
\end{align*}
where $c=(1-2p)x$.\\

Here, we give and fix a graph of $f(x)$ on $[0,a]$, that is, $y=f(x)$ for $x\in[0,a]$. Next, we introduce an evaluation function $V^{(n)}_a(p)$ for $a\in \RM_>,\  n \in \ZM_>$, and $p \in[0,1]$ with respect to $\{ \mu_x(y,p) : x \in \RM_>,\ y \in \RM \}$ in the following:
\begin{align*}
V^{(n)}_a(p) = \int_0^a dx\int_\RM dy \left(y-f(x)\right)^n\mu_x(y,p),
\end{align*}
where $\ZM_{>}=\{1,2,3,\ldots\}$. Remark that $V^{(n)}_a(p)$ is sometimes written as $V^{(W,n)}_a(p)\ $\\$\left(W \in \left\{ CTQW,\ DTQW,\ RW\right\}\right)$ in order to clarify the walk we consider. Let $M^{(n)}(x,p)$ be the $n$-th moment of $\mu_x(\cdot,p)$, that is,
\begin{align*}
M^{(n)}(x,p) = \int_\RM y^n\mu_x(y,p)dy \qquad (x \in \RM_>,\ p \in [0,1]).
\end{align*}
Then we obtain an expression of $V^{(n)}_a(p)$ by using $M^{(k)}(x,p)$ for $k=0,1,\ldots,n$ as follows.
\begin{prop}\label{ev 2m moment}
\begin{align*}
V^{(n)}_a(p) = \sum_{k=0}^{n}
\begin{pmatrix}
n\\
k
\end{pmatrix}
(-1)^{n-k} \int_0^a  f^{n-k}(x) M^{(k)}(x,p)  dx.
\end{align*}
\end{prop}
From now on, we focus on $n=2m\ (m \in \ZM_>)$ case. In particular, for $n=2$ and $n=4$ cases, Proposition \ref{ev 2m moment} gives
\begin{cor}
\begin{align}
V^{(2)}_a(p) &= \int_0^a M^{(2)}(x,p)dx -2\int_0^a f(x)M^{(1)}(x,p)dx + \int_0^a f^2(x) M^{(0)}(x,p)dx, \label{ev 2 moment}\\
V^{(4)}_a(p) &=\int_0^a M^{(4)}(x,p)dx -4\int_0^a f(x)M^{(3)}(x,p)dx \nonumber\\
&\qquad+ 6\int_0^a f^2(x) M^{(2)}(x,p)dx  - 4\int_0^a f^3(x) M^{(1)}(x,p)dx \nonumber\\
&\qquad + \int_0^a f^4(x) M^{(0)}(x,p)dx. \label{ev 4 moment}
\end{align}
\end{cor}
Next, we find all $p$'s minimizing $V_a^{(n)}(p)$. If the $p$ is uniquely obtained, then let $p_\ast^{(W, n)}(a)$ be $p$. If the $p$ is not uniquely obtained, let $p_\ast^{(W, n)}(a)$ be an average of $p$. We will discuss the expectation using this $p_\ast^{(W, n)}(a)$ in Section 7.

% 3.
%%%%%%%%%%%%%%%%%%%%%%%%%%%%%%%%%%%%%%%%%%%%%%%%%%%%%%%%%%%%%%%%%%%%%%%%%%%%%%
\section{Evaluation Function \label{sec3}} 
In this section, we compute $V_a^{(2m)}(p)$ and $\partial V_a^{(2m)}(p)/\partial p$ for three models, i.e., CTQW, DTQW, and RW (= CTRW, DTRW).
\subsection{CTQW \label{subsection CTQW 2m}}
First, we begin with
\begin{align*}
M^{(k)}(x,p) &= \int_{-1}^{1} \frac{(xu+c)^k}{\pi\sqrt{1-u^2}} du = \sum_{l=0}^{k} 
\begin{pmatrix}
k\\
l
\end{pmatrix}
x^l c^{k-l} \int_{-1}^{1} \frac{u^l }{\pi\sqrt{1-u^2}} du.
\end{align*}
Next, by the following well-known fact (see \cite{spectral analysis}, for example):
\begin{align*}
\int_{-1}^{1} \frac{u^l }{\pi\sqrt{1-u^2}} du = I_\ZM(l/2) \times
\begin{pmatrix}
l \\
l/2
\end{pmatrix}
\left( \frac{1}{2} \right)^l,
\end{align*}
we get
\begin{align}\label{CTQW moment}
M^{(k)}(x,p) %&= \sum_{l=0}^{k} I_\ZM(l/2)
%\begin{pmatrix}
%k\\
%l
%\end{pmatrix}
%x^l c^{k-l} \begin{pmatrix}
%l \\
%l/2
%\end{pmatrix}
%\left( \frac{1}{2} \right)^l  
=  c^{(CTQW)}(k,p)\ x^k,
\end{align} 
where
\begin{align}\label{CTQW c}
c^{(CTQW)}(k,p) = \sum_{l=0}^{k} I_\ZM(l/2)
\begin{pmatrix}
k\\
l
\end{pmatrix}
\begin{pmatrix}
l \\
l/2
\end{pmatrix}
(1-2p)^{k-l} \left(\frac{1}{2}\right)^l.
\end{align}
Remark that we used $c=(1-2p)x$ in order to obtain Eq. (\ref{CTQW moment}).
Combining Proposition \ref{ev 2m moment} with Eqs. (\ref{CTQW moment}) and (\ref{CTQW c}), we have

\begin{prop}[CTQW] \label{CTQW en 2m}
The evaluation function for CTQW is as follows.
\begin{align}\label{CTQW 2m formula}
V^{(2m)}_a(p) = \sum_{k=0}^{2m} 
\begin{pmatrix}
2m \\
k
\end{pmatrix}
(-1)^{2m-k}  c^{(CTQW)}(k,p) \int_0^a  x^k f^{2m-k}(x)  dx,
\end{align}
where 
\begin{align*}
c^{(CTQW)}(k,p) &= \sum_{l=0}^{k} I_\ZM(l/2)
\begin{pmatrix}
k\\
l
\end{pmatrix}
\begin{pmatrix}
l \\
l/2
\end{pmatrix}
(1-2p)^{k-l} \left(\frac{1}{2}\right)^l \qquad (k=0,1,\ldots,2m).
\end{align*}
\end{prop}
From now on, noting that parameter $p$ exists only in $c^{(CTQW)}(k,p)$ for the right-hand side of Eq. (\ref{CTQW 2m formula}), we calculate in  the following way:
\begin{align*}
\frac{\partial V^{(2m)}_a(p)}{\partial p} = \sum_{k=0}^{2m}
\begin{pmatrix}
2m \\
k
\end{pmatrix}
(-1)^{2m-k} \ \frac{\partial c^{(CTQW)}(k,p)}{\partial p} \int_0^a x^k f^{2m-k}(x)   dx.
\end{align*}
Then, we have
\begin{align*}
\frac{\partial c^{(CTQW)}(k,p)}{\partial p} &= \sum_{l=0}^{k} I_\ZM(l/2)
\begin{pmatrix}
k\\
l
\end{pmatrix}
\begin{pmatrix}
l \\
l/2
\end{pmatrix}
\left\{ \frac{\partial }{\partial p} (1-2p)^{k-l}\right\} \left(\frac{1}{2}\right)^l \\
&= -2\sum_{l=0}^{k} I_\ZM(l/2)
\begin{pmatrix}
k\\
l
\end{pmatrix}
\begin{pmatrix}
l \\
l/2
\end{pmatrix}
(k-l)(1-2p)^{k-l-1}\left(\frac{1}{2}\right)^l.
\end{align*}
Hence, we get
\begin{cor}[CTQW]\label{partial diff V CTQW 2m}
\begin{align*}
\frac{\partial V^{(2m)}_a(p)}{\partial p} =  -2&\sum_{k=0}^{2m} 
\begin{pmatrix}
2m \\
k
\end{pmatrix}
(-1)^{2m-k} \int_0^a x^k f^{2m-k}(x)  dx \\
& \quad\times \sum_{l=0}^{k} I_\ZM(l/2)
\begin{pmatrix}
k\\
l
\end{pmatrix}
\begin{pmatrix}
l \\
l/2
\end{pmatrix}
(k-l)(1-2p)^{k-l-1}\left(\frac{1}{2}\right)^l .
\end{align*}
\end{cor}

\subsection{DTQW\label{subsection DTQW 2m}}
As in the case of CTQW, we start with
\begin{align*}
M^{(k)}(x,p) &= \int_{c-rx}^{c+rx} \frac{y^k\sqrt{1-r^2}}{\pi x\left(1-\left(\frac{y-c}{x}\right)^2\right)\sqrt{r^2-\left(\frac{y-c}{x}\right)^2}} dy\\
&= \int_{-1}^{1} \frac{(xu+c)^k\sqrt{1-r^2}}{\pi\left( 1-u^2\right)\sqrt{r^2-u^2}} du \\
%&= \int_{-1}^{1} \displaystyle\sum_{l=0}^{k}
%\begin{pmatrix}
%k\\
%l
%\end{pmatrix}
%(xu)^l c^{k-l} \times \frac{\sqrt{1-r^2}}{\pi\left( 1-u^2\right)\sqrt{r^2-u^2}} du\\
&= \sum_{l=0}^{k} 
\begin{pmatrix}
k\\
l
\end{pmatrix}
x^l c^{k-l} \int_{-1}^{1} \frac{u^l \sqrt{1-r^2}}{\pi(1-u^2)\sqrt{r^2-u^2}} du.
\end{align*}
By the result given in Hamada et al. \cite{konno function moment}:
\begin{align*}
\int_{-1}^{1} \frac{u^l \sqrt{1-r^2}}{\pi(1-u^2)\sqrt{r^2-u^2}} du = I_{\ZM}\left(l/2\right) \times 
\left\{ 
1-\sqrt{1-r^2}\sum_{s=0}^{\frac{l}{2}-1} 
\begin{pmatrix}
2s \\
s
\end{pmatrix}
\left( \frac{r^2}{4} \right)^s
\right\},
\end{align*}
we have
\begin{align}\label{DTQW moment}
M^{(k)}(x,p) %&= \sum_{l=0}^{k}  I_{\ZM}\left(l/2\right)
%\begin{pmatrix}
%k\\
%l
%\end{pmatrix}
%x^l c^{k-l} 
%\left\{ 
%1-\sqrt{1-r^2}\sum_{x=0}^{\frac{l}{2}-1} 
%\begin{pmatrix}
%2s \\
%s
%\end{pmatrix}
%\left( \frac{r^2}{4} \right)^s
%\right\} \nonumber \\
&= c^{(DTQW)}(k,p)\ x^k,
\end{align}
where
\begin{align}
c^{(DTQW)}(k,p) &= \sum_{l=0}^{k}  I_{\ZM}\left(l/2\right)
\begin{pmatrix}
k\\
l
\end{pmatrix}
(1-2p)^{k-l}   A_r(l),\label{DTQW c}\\
A_r(l) &= 1-\sqrt{1-r^2}\sum_{s=0}^{\frac{l}{2}-1} 
\begin{pmatrix}
2s \\
s
\end{pmatrix}
\left( \frac{r^2}{4} \right)^s. \label{DTQW A}
\end{align}
From Proposition \ref{ev 2m moment} and Eqs. (\ref{DTQW moment}), (\ref{DTQW c}), and (\ref{DTQW A}), we get
\begin{prop}[DTQW]\label{DTQW en 2m}
The evaluation function for DTQW is as follows.
\begin{align*}
V^{(2m)}_a(p) = \sum_{k=0}^{2m}
\begin{pmatrix}
2m\\
k
\end{pmatrix}
(-1)^{2m-k} c^{(DTQW)}(k,p) \int_0^a  x^k f^{2m-k}(x)  dx,
\end{align*}
where
\begin{align*}
c^{(DTQW)}(k,p) &= \sum_{l=0}^{k}  I_{\ZM}\left(l/2\right)
\begin{pmatrix}
k\\
l
\end{pmatrix}
(1-2p)^{k-l}  A_r(l),\\
A_r(l) &= 1-\sqrt{1-r^2}\sum_{s=0}^{\frac{l}{2}-1} 
\begin{pmatrix}
2s \\
s
\end{pmatrix}
\left( \frac{r^2}{4} \right)^s.
\end{align*}
\end{prop}
As for $\displaystyle \partial V_a^{(2m)}(p)/\partial p$, we have
\begin{align*}
\frac{\partial V^{(2m)}_a(p)}{\partial p} &= \sum_{k=0}^{2m}
\begin{pmatrix}
2m\\
k
\end{pmatrix}
(-1)^{2m-k}\ \frac{c^{(DTQW)}(k,p)}{\partial p} \int_0^a  x^k f^{2m-k}(x)  dx.
\end{align*}
Then we get 
\begin{align*}
\frac{\partial c^{(DTQW)}(k,p)}{\partial p} &= \sum_{l=0}^{k}  I_{\ZM}\left(l/2\right)
\begin{pmatrix}
k\\
l
\end{pmatrix}
\left\{ \frac{\partial }{\partial p} (1-2p)^{k-l} \right\} A_r(l)\\
&= -2\sum_{l=0}^{k}  I_{\ZM}\left(l/2\right)
\begin{pmatrix}
k\\
l
\end{pmatrix}
(k-l)(1-2p)^{k-l-1} A_r(l).
\end{align*}
Therefore, we obtain
\begin{cor}[DTQW]\label{partial diff V DTQW 2m}
\begin{align*}
\frac{\partial V^{(2m)}_a(p)}{\partial p} = -2&\sum_{k=0}^{2m}
\begin{pmatrix}
2m\\
k
\end{pmatrix}
(-1)^{2m-k} \int_0^a  x^k f^{2m-k}(x)  dx \\
& \quad \times\sum_{l=0}^{k}  I_{\ZM}\left(l/2\right)
\begin{pmatrix}
k\\
l
\end{pmatrix}
(k-l)(1-2p)^{k-l-1} A_r(l).
\end{align*}
\end{cor}

\subsection{RW (= CTRW, DTRW)\label{BM 2m}}
Similarly, we have
\begin{align*}
M^{(k)}(x,p) &= \int_\RM y^k \times \frac{1}{\sqrt{2\pi x}} e^{-\frac{\left( y-c \right)^2}{2x}} dy = \int_\RM \left( \sqrt{x}u+c \right)^k \times \frac{1}{\sqrt{2\pi}} e^{-\frac{u^2}{2}} du  \\
&= \sum_{l=0}^{k} \begin{pmatrix}
k\\
l
\end{pmatrix}
x^{\frac{l}{2}} c^{k-l} \int_\RM u^l \frac{1}{\sqrt{2\pi}}e^{-\frac{u^2}{2}} du.
\end{align*}
From the following well-known fact (see \cite{spectral analysis}, for example):
\begin{align*}
\int_\RM u^l \frac{1}{\sqrt{2\pi}}e^{-\frac{u^2}{2}} du = I_\ZM\left( l/2\right) \times \frac{l!}{2^{\frac{l}{2}}\left(\frac{l}{2}\right)!},
\end{align*}
we get
\begin{align}
M^{(k)}(x,p) &= \sum_{l=0}^{k} I_\ZM\left( l/2\right)
\begin{pmatrix}
k\\
l
\end{pmatrix}
 (1-2p)^{k-l}\frac{l!}{2^{\frac{l}{2}}\left(\frac{l}{2}\right)!} x^{k-\frac{l}{2}}. \label{BM moment}
\end{align}
By Proposition \ref{ev 2m moment} and Eq. (\ref{BM moment}), we obtain
\begin{prop}[RW]\label{RW en 2m}
The evaluation function for RW is as follows.
\begin{align*}
V^{(2m)}_a(p) =  \sum_{k=0}^{2m}
\begin{pmatrix}
2m\\
k
\end{pmatrix}
(-1)^{2m-k} 
&\sum_{l=0}^{k} I_\ZM\left( l/2\right) 
\begin{pmatrix}
k\\
l
\end{pmatrix}
(1-2p)^{k-l} \frac{l!}{2^{\frac{l}{2}}\left(\frac{l}{2}\right)!} \\
& \qquad \qquad \quad \times \int_0^a x^{k-\frac{l}{2}}f^{2m-k}(x) dx.
\end{align*}
\end{prop}
As in the case of CTQW and DTQW, we have
%\ref{subsection CTQW 2m} and \ref{subsection DTQW 2m},
\begin{cor}[RW]
\begin{align*}
\frac{\partial V^{(2m)}_a(p)}{\partial p} = -2\sum_{k=0}^{2m}
\begin{pmatrix}
2m\\
k
\end{pmatrix}
(-1)^{2m-k} 
&\sum_{l=0}^{k} I_\ZM\left( l/2\right) 
\begin{pmatrix}
k\\
l
\end{pmatrix}
(k-l)(1-2p)^{k-l-1} \\
& \qquad\times \frac{l!}{2^{\frac{l}{2}}\left(\frac{l}{2}\right)!} \int_0^a x^{k-\frac{l}{2}}f^{2m-k}(x) dx.
\end{align*}
\end{cor}
%\begin{gather*}
%\sum_{x\in\ZM} x^2\sqrt{\frac{1-p}{p}}^xI_x(2\sqrt{p(1-p)}t) = \frac{2\sqrt{p(1-p)}t}{2}\\
%\left\{\left(\sqrt{\frac{1-p}{p}}+\left(\sqrt{\frac{1-p}{p}}\right)^{-1}\right)+\frac{2\sqrt{p(1-p)}t}{2}\left(\sqrt{\frac{1-p}{p}}-\left(\sqrt{\frac{1-p}{p}}\right)^{-1}\right)^2\right\}\\
%\exp\left(\frac{2\sqrt{p(1-p)}t}{2}\left(\sqrt{\frac{1-p}{p}}+\left(\sqrt{\frac{1-p}{p}}\right)^{-1}\right) \right)\\
%M_2(t) = t + (1-2p)^2t^2\\
%\end{gather*}

% 4.
%%%%%%%%%%%%%%%%%%%%%%%%%%%%%%%%%%%%%%%%%%%%%%%%%%%%%%%%%%%%%%%%%%%%%%%%%%%%%%
\section{Special Cases for $n$ \label{sec4}}
In this section, we deal with $n=2$ and $n=4$ cases for each example. Concerning the range of $p$, we consider not $[0,1]$ but $\RM$.
\subsection{CTQW \label{subsection CTQW 2 4}}
(i) \textbf{$n=2$}.
In this case, we get
\begin{align*}
M^{(0)}(x,p) = 1, \quad M^{(1)}(x,p) = (1-2p)x, \quad M^{(2)}(x,p) = \left\{ \frac{1}{2}+(1-2p)^2 \right\}x^2.
\end{align*}
By Eq. (\ref{ev 2 moment}), we have
\begin{align*}
V^{(2)}_a(p) %= &\int_0^a \left\{ 1+(1-2p)^2 \right\}x^2dx - 2\int_0^a f(x)(1-2p)xdx \\
%&\qquad+ \int_0^a f^2(x) dx\\
= & \left\{ \frac{1}{2}+(1-2p)^2 \right\}\frac{a^3}{3} - 2(1-2p)\int_0^a xf(x)dx + \int_0^a f^2(x) dx. 
\end{align*}
From now on, we introduce the following notations:
\begin{align}
\langle x^\alpha f^\beta \rangle &= \int_0^a  x^\alpha f^{\beta}(x) dx \qquad (\alpha , \beta \in \RM),\label{integ}\\
w &= 1-2p.\label{w}
\end{align}
Using Eqs. (\ref{integ}) and (\ref{w}), we obtain
\begin{align}
V^{(2)}_a(w) =  \langle f^2\rangle - 2w\langle xf\rangle + \left( \frac{1}{2}+w^2 \right)\frac{a^3}{3}. \label{CTQW en}
\end{align}
Remark that Eq. (\ref{CTQW en}) can also be derived from Proposition \ref{CTQW en 2m}. In order to find $w$ minimizing $V^{(2)}_a(w)$, we calculate
\begin{align}
\frac{\partial V^{(2)}_a(w)}{\partial w} = -2 \left( \langle xf\rangle - \frac{a^3}{3}w \right). \label{partial diff V CTQW}
\end{align}
Then $\partial V^{(2)}_a(w)/\partial w =0$ implies that the unique critical point on $\RM$ is
\begin{align*}
w = \frac{3}{a^3} \langle xf\rangle.
\end{align*}
On the other hand, from Eq. (\ref{CTQW en}), we observe that the graph of $V^{(2)}_a(w)$ is concave upward on $\RM$, since $\partial V^{(2)}_a(w)/\partial w = 2a^2/3 > 0$. Therefore, we have the unique global minimum value on $\RM$ at $w = 3 \langle xf\rangle/a^3$. Equivalently, as for $p=(1-w)/2$, we obtain the corresponding global minimum value on $\RM$ at
\begin{align}
%\frac{4}{3}(1-2p)a^3=4\int_0^a f(x)x dx\nonumber\\
%1-2p = \frac{3}{a^2}\int_0^a f(x)x dx\nonumber\\
p = \frac{1}{2} - \frac{3}{2a^3}\langle xf\rangle. \label{CTQW p}
\end{align}\\

(ii) \textbf{$n=4$}.
From Proposition \ref{CTQW en 2m}, we compute
\begin{align*}
V_a^{(4)}(p) %&= \sum_{k=0}^{4} 
%\begin{pmatrix}
%4 \\
%k
%\end{pmatrix}
%(-1)^{4-k}  c^{(CTQW)}(k,p) \int_0^a  x^k f^{4-k}(x)  dx\\
&= c^{(CTQW)}(0,p) \int_0^a    f^{4}(x)  dx
-4c^{(CTQW)}(1,p) \int_0^a  x  f^{3}(x)  dx \\
& \qquad +6c^{(CTQW)}(2,p) \int_0^a   x^2 f^{2}(x)  dx
-4c^{(CTQW)}(3,p) \int_0^a  x^3 f(x)  dx \\
&\qquad +c^{(CTQW)}(4,p) \int_0^a  x^4   dx\\
 &= \langle f^4\rangle-4w\langle xf^3\rangle+6\left(w^2+\frac{1}{2}\right)\langle x^2f^2\rangle\nonumber\\
&\qquad-4w\left(w^2+\frac{3}{2}\right)\langle x^3f\rangle+\left(w^4+3w^2+\frac{3}{8}\right)\frac{a^5}{5}.
\end{align*}
Combining the result on $n=2$ with that on $n=4$ implies
\begin{prop}[CTQW. $n=2,4$]\label{CTQW ev 2 4}
The evaluation functions of the CTQW for $n=2$ and $n=4$ cases are given by
\begin{align*}
V^{(2)}_a(w) &=  \langle f^2\rangle - 2\langle xf\rangle w + \frac{a^3}{3} \left( \frac{1}{2}+w^2 \right),\\
V_a^{(4)}(w) &= \langle f^4\rangle - 4\langle xf^3\rangle w + 6\langle x^2f^2\rangle\left(w^2+\frac{1}{2}\right)\nonumber\\
&\qquad-4\langle x^3f\rangle w\left(w^2+\frac{3}{2}\right) + \frac{a^5}{5}\left(w^4+3w^2+\frac{3}{8}\right),
\end{align*}
where $w=1-2p$ for $p\in\RM$. For $n=2$ case, $V^{(2)}_a(p)$ has the unique global minimum on $\RM$ at
\begin{align*}
p = \frac{1}{2} - \frac{3}{2a^3}\langle xf\rangle.
\end{align*}
\end{prop}
\subsection{DTQW\label{subsection DTQW 2 4}}
(i) \textbf{$n=2$}.
In this case, we get
\begin{align*}
M^{(0)}(x,p) &= 1, \qquad M^{(1)}(x,p) = (1-2p)x, \\
M^{(2)}(x,p) &= \left\{ 1-\sqrt{1-r^2}+(1-2p)^2 \right\}x^2.
\end{align*}
As in the case of CTQW, we calculate
\begin{align}
V^{(2)}_a(w) &= \langle f^2\rangle - 2w\langle xf\rangle + \left( 1-\sqrt{1-r^2}+w^2 \right)\frac{a^3}{3}, \label{DTQW en}\\
\frac{\partial V^{(2)}_a(w)}{\partial w} &= -2 \left( \langle xf\rangle - \frac{a^3}{3}w \right). \label{partial diff V DTQW}
\end{align}
Remark that Eq. (\ref{DTQW en}) can also be derived from Proposition \ref{CTQW en 2m}. A similar argument on CTQW implies that $V_a^{(2)}$ has the unique global minimum on $\RM$ at
\begin{align*}
w = \frac{3}{a^3} \langle xf\rangle.
\end{align*}
%\begin{align}
%p = \frac{1}{2} - \frac{3}{2a^3}\int_0^a xf(x) dx\label{DTQW p}.
%\end{align}

(ii) \textbf{$n=4$}.
As in the case of CTQW for $n=4$, Proposition \ref{CTQW en 2m} gives
\begin{align*}
V_a^{(4)}(p) &= c^{(DTQW)}(0,p)\langle f^4\rangle - 4c^{(DTQW)}(1,p)\langle xf^3\rangle + 6c^{(DTQW)}(2,p)\langle x^2f^2\rangle \\
& \qquad - 4c^{(DTQW)}(3,p)\langle x^3f\rangle + c^{(DTQW)}(4,p)\langle x^4\rangle \\
&= \langle f^4\rangle - 4w\langle xf^3\rangle + 6\left( w^2+1-\sqrt{1-r^2}\right)\langle x^2f^2\rangle \\
& \qquad - 4\left\{ w^3+3(1-\sqrt{1-r^2})w\right\}\langle x^3f\rangle \\
& \qquad + \left\{w^4 + 6(1-\sqrt{1-r^2})w^2 + 1-\sqrt{1-r^2}\left(1+\frac{r^2}{2}\right)\right\}\frac{a^5}{5}.
\end{align*}
Combining the result on $n=2$ with that on $n=4$ implies
\begin{prop}[DTQW. $n=2,4$]\label{DTQW ev 2 4}
The evaluation functions of the DTQW for $n=2$ and $n=4$ cases are given by
\begin{align*}
V^{(2)}_a(w) &= \langle f^2\rangle - 2\langle xf\rangle w + \frac{a^3}{3}\left( 1-\sqrt{1-r^2}+w^2 \right), \\
V^{(4)}_a(w) &= \langle f^4\rangle - 4\langle xf^3\rangle w + 6\langle x^2f^2\rangle\left( w^2+1-\sqrt{1-r^2}\right) \\
& \qquad - 4\langle x^3f\rangle\left\{ w^3+3(1-\sqrt{1-r^2})w\right\} \\
& \qquad + \frac{a^5}{5}\left\{w^4 + 6(1-\sqrt{1-r^2})w^2 + 1-\sqrt{1-r^2}\left(1+\frac{r^2}{2}\right)\right\}. 
\end{align*}
where $w=1-2p$ for $p\in\RM$. For $n=2$ case, $V^{(2)}_a(p)$ has the unique global minimum on $\RM$ at
\begin{align*}
p = \frac{1}{2} - \frac{3}{2a^3}\langle xf\rangle.
\end{align*}
\end{prop}

\subsection{RW (= CTRW, DTRW)}
(i) \textbf{$n=2$}.
In this case, we get
\begin{align*}
M^{(0)}(x,p) &= 1, \qquad M^{(1)}(x,p) = (1-2p)x, \\
M^{(2)}(x,p) &= x+(1-2p)^2x^2.
\end{align*}
As in the case of CTQW, we compute
\begin{align}
V^{(2)}_a(w) =& \langle f^2\rangle  - 2w\langle xf\rangle +w^2\frac{a^3}{3} + \frac{a^2}{2}, \label{bm ev}\\
\frac{\partial V^{(2)}_a(w)}{\partial w} &= -2 \left( \langle xf\rangle - \frac{a^3}{3}w \right). \label{partial diff V bm}
\end{align}
Remark that Eq. (\ref{bm ev}) can also be derived from Proposition \ref{CTQW en 2m}. A similar argument on CTQW and DTQW implies that $V^{(2)}_a(w)$ has the unique global minimum on $\RM$ at
\begin{align*}
p = \frac{1}{2} - \frac{3}{2a^3}\langle xf\rangle.
\end{align*}
Note that Eqs. (\ref{partial diff V CTQW}), (\ref{partial diff V DTQW}), and (\ref{partial diff V bm}) are the same.

(ii) \textbf{$n=4$}.
As in the case of CTQW and DTQW for $n=4$, from Proposition \ref{CTQW en 2m}, we have
\begin{align*}
V_a^{(4)}(w) &= \langle f^4\rangle - 4w\langle xf^3\rangle + 6\left( w^2\langle x^2f^2\rangle + \langle xf^2\rangle\right)\\
& \qquad - 4\left( w^3\langle x^3f\rangle+3w\langle x^2f\rangle\right) + \left( w^4\frac{a^5}{5}+6w^2\frac{a^4}{4}+3\frac{a^3}{3}\right).
\end{align*}
Combining the result on $n=2$ with that on $n=4$ implies
\begin{prop}[RW. $n=2,4$]\label{RW ev 2 4}
The evaluation functions of the RW for $n=2$ and $n=4$ cases are given by
\begin{align*}
V^{(2)}_a(w) &= \langle f^2\rangle  - 2\langle xf\rangle w +\frac{a^3w^2}{3} + \frac{a^2}{2},\\
V_a^{(4)}(w) &= \langle f^4\rangle - 4\langle xf^3\rangle w + 6\left( \langle x^2f^2\rangle w^2 + \langle xf^2\rangle\right)\\
& \qquad - 4\left( \langle x^3f\rangle w^3 + 3\langle x^2f\rangle w \right) + \left( \frac{a^5w^4}{5}+\frac{3a^4w^2}{2}+a^3\right).
\end{align*}
For $n=2$ case, $V^{(2)}_a(p)$ has the unique global minimum on $\RM$ at
\begin{align*}
p = \frac{1}{2} - \frac{3}{2a^3}\langle xf\rangle.
%w = \frac{3}{a^3} \langle xf\rangle.
\end{align*}
\end{prop}
We should note that $V^{(2)}_a(p)$ for each CTQW, DTQW, and RW case has the unique global minimum value on $\RM$ at the same
\begin{align*}
p = \frac{1}{2} - \frac{3}{2a^3}\langle xf\rangle.
\end{align*}

Remark that the variance of each walk doesn't depend on $p$. See below for detail:
\begin{align}
\sigma^{2,(CTQW)} &= \left\{ \frac{1}{2}+(1-2p)^2 \right\}x^2 - \left\{ (1-2p)x \right\}^2\nonumber\\
&= \frac{x^2}{2}, \label{variance ctqw}\\
\sigma^{2,(DTQW)} &= \left\{ 1-\sqrt{1-r^2}+(1-2p)^2 \right\}x^2 - \left\{ (1-2p)x \right\}^2\nonumber\\
&= \left(1-\sqrt{1-r^2}\right)x^2, \label{variance dtqw}\\
\sigma^{2,(RW)} &= x+(1-2p)^2x^2 - \left\{ (1-2p)x \right\}^2\nonumber\\
&= x, \label{variance rw}
\end{align}
where $\sigma^{2,(W)}\ (W \in\left\{ CTQW,\ DTQW,\ RW\right\})$ is the variance of each walk.
% 5.
%%%%%%%%%%%%%%%%%%%%%%%%%%%%%%%%%%%%%%%%%%%%%%%%%%%%%%%%%%%%%%%%%%%%%%%%%%%%%%%
\section{Special Cases for $f(x)$ \label{sec5}}
In this section, we deal with $f(x)=x$ and $f(x)=\cos x$ cases for CTQW, DTQW, and RW. As in the previous section, concerning the range of $p$, we treat not $[0,1]$ but $\RM$.\\
(A) CTQW.\\
(a) $f(x)=x$ case.\\
  (i) $n=2$.
Noting that $\langle xf\rangle = a^3/3$, Eq. (\ref{CTQW p}) gives
\begin{align}
p &= \frac{1}{2} - \frac{3}{2a^3}\times\frac{a^3}{3} = 0. \label{p=0}
\end{align}
From Eq. (\ref{mean}), we see that $p$ can be interpreted as the probability that the walker we consider moves to the left (that is, negative direction). On the other hand, when $f(x)=x$, the trajectory always moves to the right (that is, positive direction). Thus, we can expect $p=0$, which is consistent with Eq. (\ref{p=0}). \\
  (ii) $n=4$.
First, we observe that
\begin{align}
\langle f^4\rangle=\langle xf^3\rangle=\langle x^2f^2\rangle=\langle x^3f\rangle=\frac{a^5}{5}.\label{x integ}
\end{align}
Thus, by Eq. (\ref{x integ}) and Proposition \ref{CTQW ev 2 4}, we have
\begin{align*}
V^{(4)}_a(w) &= \frac{a^5}{5}\left(w^4-4w^3+9w^2-10w+\frac{35}{8}\right),\\
\frac{\partial V^{(4)}_a(w)}{\partial w} &= \frac{2a^5}{5} h_1(w),
\end{align*}
where $h_1(w) =  2w^3-6w^2 + 9w -5$. Here, we focus on $w\in[-1,1]$, that is, $p\in[0,1]$.
Since $h_1(-1)=-22<0$, $h_1(1)=0$ and $h'_1(w)= 6(w-1)^2+3 >0$, we get $\partial V^{(2)}_a(w)/\partial w \leq 0$ for $w\in [-1,1]$. Therefore, we see that $V^{(4)}_a(w)$ is decreasing on $[-1,1]$ and $V^{(4)}_a(w)$ has the unique global minimum value on $[-1,1]$ at $w=1$, that is, at $p=0$. This is the same as that for $n=2$ case.\\
(b) $f(x)=\cos x$ case.\\
  (i) $n=2$.
$\langle xf\rangle = a \sin a +\cos a - 1$ and Eq. (\ref{CTQW p}) implies that $V_a^{(2)}$ has the unique minimum value on $\RM$ at
\begin{align*}
p = \frac{1}{2} - \frac{3}{2a^3}\left( a \sin a +\cos a - 1\right).
\end{align*}\\
(ii) $n=4$.
In this case, we have
\begin{align}
\langle f^4\rangle &=\frac{1}{32} \left( 12a + 8\sin 2a + \sin 4a \right), \label{CTQW cos 1}\\
\qquad \langle x f^3\rangle &= \frac{1}{36}\left( 27a\sin a + 3a\sin 3a + 27\cos a + \cos 3a - 28 \right) ,\label{CTQW cos 2}\\
\qquad \langle x^2f^2\rangle &= \frac{1}{24}\left\{ 4a^3 + (6a^2-3)\sin 2a + 6a\cos 2a\right\}, \label{CTQW cos 3}\\
\qquad \langle x^3f\rangle &= a(a^2-6)\sin a + 3(a^2-2)\cos a + 6.\label{CTQW cos 4}
\end{align}
Thus, it follows from Eqs. (\ref{CTQW cos 1}), (\ref{CTQW cos 2}), (\ref{CTQW cos 3}), (\ref{CTQW cos 4}), and Proposition \ref{CTQW ev 2 4} that we see
\begin{align*}
V_a^{(4)} &= A_4w^4 + A_3w^3 + A_2 w^2 + A_1w + A_0,
% &= \frac{a^5}{5}w^4 + \left( -4a^3\sin a + 24a\sin a - 12 a^2 \cos  a + 24 \cos a -24\right)w^3 \\&&\qquad + \left( \frac{3}{5}a^5 + \frac{3}{2}a^3\sin 2a -\frac{3}{4} \sin 2a + \frac{3}{2}a\cos 2a + a^3\right)w^2\\
%&\qquad + \left( -\frac{1}{3}a\sin 3a - \frac{1}{9}\cos 3a - 6a^3\sin a + 33a\sin a - 18 a^2\cos a + 33 cos a - \frac{296}{9}\right)w\\
%&\qquad + \left( \frac{3}{40}a^5 + \frac{1}{32}\sin 4a + \frac{3}{4}a^3\sin 2a - \frac{1}{3}\sin 2a + \frac{3}{4}a\cos 2a + \frac{a^3}{2} + \frac{3}{8}a\right)\\
\end{align*}
where 
\begin{align*}
A_4 &= \frac{a^5}{5}, \qquad A_3 = -4\left\{a\left(a^2 - 6\right) \sin a + 3 \left(a^2 - 2\right) \cos a + 6\right\},\\
A_2 &= \frac{3}{4}\left(2a^{2} -1 \right) \sin 2a + \frac{3}{2}a\cos 2a + \frac{3}{5}a^5 + a^3, \\
A_1 &= -\frac{1}{3}a\sin 3a - \frac{1}{9}\cos 3a - 3a\left(2a^2 - 11\right) \sin a - 3\left(6 a^2 - 11\right) \cos a - \frac{296}{9},\\
A_0 &=  \frac{1}{32}\sin 4a + \frac{1}{8} \left( 6 a^3 - 1\right) \sin 2a + \frac{3}{4}a\cos 2a + \frac{3}{40}a^5 + \frac{a^3}{2} + \frac{3}{8}a.
\end{align*}
Calculating $\partial V^{(4)}_a(w)/\partial w$ and $\partial^2 V^{(4)}_a(w)/\partial w^2$, we get
\begin{align*}
\frac{\partial V^{(4)}_a(w)}{\partial w} &= 4A_4w^3 + 3A_3 w^2 + 2A_2w + A_1,\\
\frac{\partial^2 V^{(4)}_a(w)}{\partial w^2} &= 2\left( 6A_4w^2 + 3A_3w + A_2\right).
%&= h_1(w)
\end{align*}
The discriminant $D$ of $ 6A_4w^2 + 3A_3w + A_2 = 0$ is 
\begin{align*}
D = 9A_3^2 - 24A_2A_4.
\end{align*}\par
If $a=2\pi$, then we have $A_4>0$ and
\begin{align*}
D &= 9\times 48^2 \pi^4 - \frac{24\times 32}{5}\left( \frac{96}{5}\pi^4 + 8\pi^2 + 3 \right) \pi^6 < 0. 
%&<2^8\times 3^2\times\left( 3^2 - \frac{2^5}{5^2}\pi^6\right)\pi^4 
\end{align*}
Thus, $\partial V^{(4)}_a(w)/\partial w$ is increasing and has a unique root. Since
\begin{align*}
\left. \frac{\partial V^{(4)}_a(w)}{\partial w} \right|_{w=-1} &= -4A_4 + 3A_3 - 2A_2 + A_1\\
&= -64\pi^5 - 16\pi^3 - 216\pi^2 - 6\pi < 0,\\
\left. \frac{\partial V^{(4)}_a(w)}{\partial w} \right|_{w=0} &= A_1 = -72\pi^2 < 0,\\
\left. \frac{\partial V^{(4)}_a(w)}{\partial w} \right|_{w=1} &= 4A_4 + 3A_3 + 2A_2 + A_1\\
&= 64\pi^5 + 16\pi^3 - 216\pi^2 + 6\pi > 0,
\end{align*}
we see that a root of $\partial V^{(4)}_a(w)/\partial w$ exists on $(0,1)$. Therefore, $V^{(4)}_a(w)$ has the unique global minimum value on $(0,1)$. \par
In a similar way, if $a=\pi$, we see that $V^{(4)}_a(w)$ has the unique global minimum value on $(-1,0)$.

(B) DTQW.\\
(a) $f(x)=x$ case.\\
(i) $n=2$.
The argument for DTQW is almost the same as that for CTQW. Therefore we have the same conclusion.\\
%Similar to example 1, we get
%\begin{align*}
%p &= \frac{1}{2} - \frac{3}{2a^3}\times\frac{a^3}{3} = 0,
%\end{align*}
%となり，直観に沿った結果を得る．\\
%\color{blue}
%which is consistent with intuition.
(ii) $n=4$.
From Eq. (\ref{x integ}) and Proposition \ref{DTQW ev 2 4}, we get
\begin{align*}
V_a^{(4)}(w) &= \frac{a^5}{5}\left\{ w^4 - 4w^3 + 6(1+B(r))w^2 - 4\left(1+3B(r)\right)w + 1 \right.\\
& \qquad\qquad\qquad\qquad\qquad\qquad\qquad\qquad \left.  + 7B(r) - \sqrt{1-r^2}\frac{r^2}{2}\right\},\\
\frac{\partial V_a^{(4)}(w)}{\partial w} %&= \frac{4a^5}{5}\left\{ w^3 - 3w^2 + 3(1+B(r))w - \left(1+3B(r)\right) \right\}\\
&= \frac{4a^5}{5}h_2(w),
\end{align*}
where $B(r) = 1-\sqrt{1-r^2}(>\!\!0)$ and $h_2(w) = w^3 - 3w^2 + 3(1+B(r))w - \left(1+3B(r)\right)$. From
\begin{align*}
h_2'(w) %&= 3 w^2 - 6w + 3(1+B(r)) 
=  3\left\{ (w-1)^2 + B(r)\right\} > 0,
\end{align*}
$h_2(w)$ is increasing. Thus, $h_2(1)=0$ gives $\partial V_a^{(4)}(w)/\partial w\leq0$ on $[-1,1]$. Therefore $V_a^{(4)}(p)$ has the global minimum value on [0,1] at $p=0$.\\
(b) $f(x)=\cos x$ case.\\
(i) $n=2$. 
This case is almost the same as CTQW case.\\
(ii) $n=4$. As in the case of CTQW, we have
\begin{align*}
V_a^{(4)}(p) &= A_4w^4 + A_3w^3 + A_2 w^2 + A_1w + A_0,
\end{align*}
where
\begin{align*}
A_4 &= \frac{a^5}{5}, \qquad A_3 =  -4\left\{a\left(a^2 - 6\right) \sin a + 3 \left(a^2 - 2\right) \cos a + 6\right\}, \\
A_2 &= \frac{3}{4}\left(2a^{2} -1 \right) \sin 2a + \frac{3}{2}a\cos 2a - \frac{6}{5}a^5\sqrt{1-r^2} + \frac{6}{5}a^5 + a^3, \\
A_1 &= 12a \left( a^2  - 6 \right)\sqrt{1-r^2} \sin a + 36\left( a^2 - 2\right)\sqrt{1-r^2}\cos a \\
&\qquad + 72\sqrt{1-r^2} - \frac{a}{3}\sin 3a - \frac{1}{9}\cos 3a - 3a\left(4a^2 - 23\right)\sin a \\
&\qquad - 3\left(12a^2- 23\right)\cos a - \frac{620}{9}, \\
A_0 &=  \frac{1}{32}\sin 4a + \left\{ -\frac{3}{4}\left(2a^2 - 1\right)\sqrt{1-r^2} + \frac{1}{2}\left( 3a^2  - 1\right)\right\} \sin 2a  \\
&\qquad + \frac{3}{2}a\left( 1 - \sqrt{1-r^2} \right) \cos 2a - \frac{a^5}{10}\left( r^2\sqrt{1-r^2} + 2\sqrt{1-r^2} - 2\right) \\
&\qquad + a^3\left( 1 - \sqrt{1-r^2} \right) + \frac{3}{8}a.
\end{align*}\par
If $a=2\pi$, then as in the case of CTQW, we get $A_4>0$ and 
\begin{align*}
D = 9\times 48^2 \pi^4 - \frac{24\times 32}{5}\left\{ \frac{6\times 32}{5}B(r)\pi^5 + 2\pi\left(\frac{3}{2} + 4\pi^2\right) \right\} \pi^5 < 0.
\end{align*}
Thus, $\partial V^{(4)}_a(w)/\partial w$ is increasing and has a unique root. Since
\begin{align*}
\left. \frac{\partial V^{(4)}_a(w)}{\partial w} \right|_{w=-1} &= -\frac{128}{5}\pi^5 - 144\pi^2 - B(r)\times \frac{2^4 \times 3}{5}\pi^2 \left\{(2\pi)^3 + 3\times 5 \right\}  \\
&\qquad\qquad\qquad\qquad\qquad\qquad\qquad - 2\pi\left\{ 2(2\pi)^2 + 3 \right\} < 0, \\
\left. \frac{\partial V^{(4)}_a(w)}{\partial w} \right|_{w=0} &= -144B(r)\pi^2 < 0,\\
\left. \frac{\partial V^{(4)}_a(w)}{\partial w} \right|_{w=1} &= \frac{128}{5}\pi^5 - 144\pi^2 - B(r)\times \frac{2^4 \times 3}{5}\pi^2 \left\{-(2\pi)^3 + 3\times 5 \right\}\\
&\qquad\qquad\qquad\qquad\qquad\qquad\qquad + 2\pi\left\{ 2(2\pi)^2 + 3 \right\} > 0,
\end{align*}
we obtain the same conclusion as in CTQW.\par
Similarly, if $a=\pi$, we observe that $V^{(4)}_a(w)$ has the unique global minimum value on $(-1,0)$.

(C) RW (= CTRW, DTRW).\\
(a) $f(x)=x$ case.\\
(i) $n=2$.
The argument for RW is almost the same as that for CTQW and DTQW. Therefore we have the same conclusion.\\
(ii) $n=4$.
It follows from $\langle f^4\rangle = \langle xf^3 \rangle = \langle x^2f^2 \rangle = \langle x^3f\rangle = a^5/5$, $\langle xf^2\rangle = \langle x^2f\rangle = a^4/4$, Eq. (\ref{x integ}) and Proposition \ref{RW ev 2 4} that we see
\begin{align*}
V_a^{(4)}(w) &=  \frac{a^5}{5}w^4 - \frac{4a^5}{5}w^3 + 6\left( \frac{a^5}{5}+\frac{a^4}{4}\right)w^2 \\
& \qquad- 4\left( \frac{a^5}{5}+\frac{3a^4}{4} \right)w + \frac{a^5}{5} + \frac{3a^4}{2} + a^3,\\
\frac{\partial V_a^{(4)}(w)}{\partial w} %&= 4\left\{ \frac{a^5}{5}w^3 - 3\frac{a^5}{5}w^2 + 3\left( \frac{a^5}{5}+\frac{a^4}{4}\right)w - \left( \frac{a^5}{5}+3\frac{a^4}{4} \right)\right\}\\
&= 4h_3(w),
\end{align*}
where
\begin{align*}
h_3(w) =  \frac{a^5}{5}w^3 - \frac{3a^5}{5}w^2 + 3\left( \frac{a^5}{5}+\frac{a^4}{4}\right)w - \left( \frac{a^5}{5}+\frac{3a^4}{4} \right).
\end{align*}
Since
\begin{align*}
h'_3(w) = \frac{3a^5}{5}(w-1)^2 + \frac{3a^4}{4} \geq 0,
\end{align*}
$h_3(w)$ is increasing. Thus, $h_3(1)=0$ gives $\partial V_a^{(4)}(w)/\partial w \leq 0$ on $[-1,1]$. That implies $V_a(p)$ has the global minimum value on $[0,1]$ at $p=0$.

(b) $f(x)=\cos x$ case.\\ 
(i) $n=2$.
This case is almost the same as CTQW and DTQW.\\
(ii) $n=4$.
For this case, from Eqs. (\ref{CTQW cos 1}), (\ref{CTQW cos 2}), (\ref{CTQW cos 3}), (\ref{CTQW cos 4}), and Proposition (\ref{RW ev 2 4}), we have
\begin{align*}
V_a^{(4)}(w) &= A_4w^4 + A_3w^3 + A_2 w^2 + A_1w + A_0,
\end{align*}
where
\begin{align*}
A_4 &= \frac{a^5}{5}, \qquad A_3 = -4\left\{a\left(a^2 - 6\right) \sin a + 3 \left(a^2 - 2\right) \cos a + 6\right\},\\
A_2 &= \frac{3}{4}\left(2a^{2} -1 \right) \sin 2a + \frac{3}{2}a\cos 2a + \frac{3}{2}a^4 + a^3,\\
A_1 &= -\frac{a}{3}\sin 3a - \frac{1}{9}\cos 3a - 3\left( 4a^2 + a - 8\right) \sin a - 3\left(8a + 1\right) \cos a \\
&\qquad + \frac{28}{9},\\
A_0 &= \frac{1}{32}\sin 4a + \frac{1}{4}\left( 6a + 1\right) \sin 2a + \frac{3}{4}\cos 2a + a^3 + \frac{3}{2}a^2 + \frac{3}{8}a - \frac{3}{4}.
\end{align*}\par
If $a=2\pi$, then as in the case of CTQW and DTQW, we have $A_4>0$ and 
\begin{align*}
D = 9\times 48^2 \pi^4 - \frac{24\times 32}{5}\left( 24\pi^3 + 8\pi^2 + 3 \right) \pi^6 < 0.
\end{align*}
we get
\begin{align*}
\left. \frac{\partial V^{(4)}_a(w)}{\partial w} \right|_{w=-1} &= -\frac{128}{5}\pi^5 - 48\pi^4 - 16\pi^3 - 144\pi^2 - 54\pi < 0,\\
\left. \frac{\partial V^{(4)}_a(w)}{\partial w} \right|_{w=0} &= -48\pi^2 < 0,\\
\left. \frac{\partial V^{(4)}_a(w)}{\partial w} \right|_{w=1} &= \frac{128}{5}\pi^5 + 48\pi^4 + 16\pi^3 - 144\pi^2 - 42\pi > 0,
\end{align*}
so the same conclusion is obtained as in CTQW and DTQW.\par
In a similar fashion, if $a=\pi$, we see that $V^{(4)}_a(w)$ has the unique global minimum value on $(-1,0)$.
% 6.
%%%%%%%%%%%%%%%%%%%%%%%%%%%%%%%%%%%%%%%%%%%%%%%%%%%%%%%%%%%%%%%%%%%%%%%%%%%%%%%
\section{Relation Between Discrete-Space Model and Our Model \label{sec6}}
The models we have studied are based on weak limit theorems for CTQW, DTQW, CTRW, and DTRW on $\ZM$. Here we deal with models based on the original CTRW and DTRW on $\ZM$. In addition, we discuss the relation between both models.\\

\textbf{Example 1}.\ \ CTRW  on $\ZM$ \\%= continuous-time/discrete-space
For this model, we consider the set of probability measures $\{\mu_x(y,p)\}$ on $\ZM$ with parameter $p\in(0,1)$ at each $x\in\RM_>$ as follows:
\begin{align*}
\mu_x(y,p) &= e^{-x}\left(\frac{1-p}{p}\right)^{\frac{y}{2}}I_y\left(2\sqrt{p(1-p)}x\right) \qquad (y\in\ZM),
\end{align*}
where $I_y(z)$ is the modified Bessel function (see \cite{special function}). Note that it is easily confirmed that $\mu_x(y,p)$ satisfies
\begin{align*}
\frac{\partial \mu_x(y,p)}{\partial x} = p\mu_x(y+1,p) + (1-p)\mu_x(y-1,p) - \mu_x(y,p).
\end{align*}
This means $\mu_x(y,p)$ can be seen as the probability distribution for an asymmetric CTRW at time $x\in\RM_>$ and at position $y\in\ZM$. Then we have
\begin{gather*}
M^{(0)}(x,p) = 1, \quad M^{(1)}(x,p) = (1-2p)x, \quad M^{(2)}(x,p) = x+(1-2p)^2x^2.
\end{gather*}
Therefore, the same conclusion for $n=2$ is obtained as in RW in Section \ref{sec4}.\\

\textbf{Example 2}.\ \ DTRW  on $\ZM$ \\%=discrete-time/discrete-space
For this case, we consider the set of probability measures $\{\mu_x(y,p)\}$ on $\ZM$ with parameter $p\in(0,1)$ at each $x\in\ZM_>$ as follows:
\begin{align*}
\mu_x(y,p) = I_\ZM\left((x-y)/2\right)
\begin{pmatrix}
x\\
(x-y)/2
\end{pmatrix}p^{\frac{x-y}{2}}(1-p)^{\frac{x+y}{2}}.
\end{align*}
Remark that $\mu_x(y,p)$ satisfies
\begin{align*}
\mu_{x+1}(y,p) = p\mu_x(y+1,p) + (1-p)\mu_x(y-1,p).
\end{align*}
This means $\mu_x(y,p)$ can be seen as the probability distribution for an asymmetric DTRW at time $x\in\ZM_>$ and at position $y\in\ZM$. Then we get
\begin{align*}
M^{(0)}(x,p) &= 1, \qquad M^{(1)}(x,p) = (1-2p)x, \\
M^{(2)}(x,p) &= 4p(1-p)x+(1-2p)^2x^2.
\end{align*}
From Proposition 4.1 in Konno \cite{TSA green}, we have
\begin{prop}\label{ev DTRW discrete time}
\begin{align*}
V_a^{(2)}(w) &= \langle f^2\rangle -2w\langle xf\rangle + \frac{a(a+1)}{6}\left\{2(a-1)w^2+3\right\}, \\
\frac{\partial V_a^{(2)}(w)}{\partial w} &= -2\left(\langle xf \rangle - \frac{(a-1)a(a+1)}{3}w \right),
\end{align*}
where $w=1-2p$ and $\displaystyle \langle x^\alpha f^\beta \rangle = \sum_{x=0}^a x^\alpha f^\beta(x)$.
\end{prop}
By Proposition \ref{ev DTRW discrete time}, we see that $V_a^{(2)}(w)$ has the unique global minimum on $\RM$ at
\begin{align*}
w = \frac{3}{(a-1)a(a+1)}\langle xf \rangle,
\end{align*}
that is, at
\begin{align}
p = \frac{1}{2} - \frac{3}{2(a-1)a(a+1)}\langle xf \rangle. \label{p compare 1}
\end{align}
Thus, if $\langle xf \rangle < \infty$, then $\displaystyle \lim_{a \to \infty}p = 1/2$.
%・$f(x)=x$の場合\\
%\color{blue}
%・$f(x)=x$ case
%\color{black}
%$\sum_{x=0}^a f(x)x = \frac{a(a+1)(2a+1)}{6}$なので，$V_a(p)$の変曲点$p$は，
%\color{blue}
%Since $\sum_{x=0}^a f(x)x = \frac{a(a+1)(2a+1)}{6}$, the inflection point $p$ of $V_a(p)$ is
%\color{black}
%\begin{align*}
%p &= \frac{1}{2} - \frac{3}{2a(a^2-1)}\times\frac{a(a+1)(2a+1)}{6} \\
%&= \frac{1}{2} - \frac{2a+1}{4(a-1)} \\
%&= \frac{2(a-1)-(2a+1)}{4(a-1)}\\
%p &= -\frac{3}{4(a-1)} < 0 \qquad (a>1)
%\end{align*}
%となる．
%よって，$p\in[0,1]$より，$p^\ast=0$が導かれる．
%\textcolor{blue}{Therefore, $p\in[0,1]$ results in $p^\ast=0$.}

On the other hand, as for our model for CTQW, DTQW, and RW, we obtained
\begin{align}
p = \frac{1}{2} - \frac{3}{2a^3}\langle xf\rangle. \label{p compare 2}
\end{align}
In this model also, if $\langle xf \rangle < \infty$, then $\displaystyle \lim_{a \to \infty}p = 1/2 $.
Here, the only difference between Eqs. (\ref{p compare 1}) and (\ref{p compare 2}) is the denominator of the second term of the right-hand side, that is, $2(a-1)a(a+1)$ and $2a^3$.
\section{Linear Extrapolation}
Put 
\begin{align*}
\mathcal M _a = \left\{ p \in (0,1)\ |\ p \text{ is a local minimum of } V_a^{\left(W,n\right)}(p)\right\}
\end{align*}
for $W \in \left\{ CTQW,\ DTQW,\ RW\right\}$, $n \in \ZM_>,$ and $a>0$. Let $p_\ast^{(W, n)}(a)$ denote that 
\begin{align*}
p_\ast^{(W, n)}(a) = \underset{p \in S}{\text{argmin}} \ V_a^{(W,n)}(p),
\end{align*}
where $S = \{0, 1, \mathcal M _a\}$ and ``argmin" means ``argument of the minimum". Then we define two linear extrapolations at location $b$ by taking expectation as follows:
\begin{align*}
m^{(W,n)}(a,b) &= \int_\RM y\ \mu_b\left(y, p_\ast^{(W, n)}(a) \right) dy, \\
\widetilde{m}^{(W,n)}(a,b) &= f(a) + \int_\RM y\ \mu_{b-a}\left(y, p_\ast^{(W, n)}(a) \right) dy,
\end{align*}
for $W \in \left\{ CTQW,\ DTQW,\ RW\right\}$, $n \in \ZM_>,$ and $b>a>0$. From Eq. (\ref{mean}), we obtain
\begin{prop}\label{expectation}
\begin{align}
m^{(W,n)}(a,b) &= \left(1 - 2 p_\ast^{(W, n)}(a) \right) b, \label{m}\\
\widetilde{m}^{(W,n)}(a,b) &= f(a) + \left(1 - 2 p_\ast^{(W, n)}(a) \right)(b-a), \label{m tilde}
\end{align}
for $W \in \left\{ CTQW,\ DTQW,\ RW\right\}$, $n \in \ZM_>,$ and $b>a>0$. 
\end{prop}
Remark that when $n \to 2$, $a \to n$, and $b \to n+1$, we see that Eqs. (\ref{m}) and (\ref{m tilde}) correspond to Eqs. (4.13) and (4.14) in Konno \cite{TSA green}, respectively. 

From now on, we consider $m^{(W,n)}(a,b)$ and $\widetilde{m}^{(W,n)}(a,b)$ for $f(x)=x$ and $f(x)=\cos x$ cases with $n=2$ and $n=4$.\\

(a)\ $f(x)=x$ case\\
From the argument for $n=2$ and $4$ in Section 5, we get
\begin{align*}
p_\ast^{(CTQW,n)}(a) = p_\ast^{(DTQW,n)}(a) = p_\ast^{(RW,n)}(a) = 0 \qquad (n=2,4).
\end{align*}
Therefore, Proposition \ref{expectation} gives,
\begin{align*}
m^{(CTQW,n)}(a,b) &= m^{(DTQW,n)}(a,b) = m^{(RW,n)}(a,b) = b, \\
\widetilde{m}^{(CTQW,n)}(a,b) &= \widetilde{m}^{(DTQW,n)}(a,b) = \widetilde{m}^{(RW,n)}(a,b) = b,
\end{align*}
for $n=2, 4$.\\

(b)\ $f(x)=\cos x$ case\\
By the argument for $n=2$ in Section 5, we have
\begin{align*}
p_\ast^{(CTQW,2)}(a) = p_\ast^{(DTQW,2)}(a) = p_\ast^{(RW,2)}(a) = \frac{1}{2} - \frac{3}{2a^3}\left( a \sin a +\cos a - 1\right).
\end{align*}
It follows from Proposition \ref{expectation} that we obtain
\begin{align*}
m^{(CTQW,2)}(a,b) &= m^{(DTQW,2)}(a,b) = m^{(RW,2)}(a,b) = \frac{3b}{a^3}\left( a \sin a +\cos a - 1\right),\\
\widetilde{m}^{(CTQW,2)}(a,b) &= \widetilde{m}^{(DTQW,2)}(a,b) = \widetilde{m}^{(RW,2)}(a,b) \\
& \quad\qquad\qquad\qquad\qquad\qquad= \cos a + \frac{3(b-a)}{a^3}\left( a \sin a +\cos a - 1\right).
\end{align*}
For $n=4$, we get
\begin{align*}
p_\ast^{(CTQW,4)}(\pi) ,\ \ p_\ast^{(DTQW,4)}(\pi) ,\ \ p_\ast^{(RW,4)}(\pi) \in \left( \frac{1}{2}, 1\right),\\
p_\ast^{(CTQW,4)}(2\pi),\ \ p_\ast^{(DTQW,4)}(2\pi),\ \ p_\ast^{(RW,4)}(2\pi) \in \left(0, \frac{1}{2}\right).
\end{align*}
Here, we assume $r=1/\sqrt{2}$ (i.e., Hadamard walk) for DTQW. Then numerical calculations show that
\begin{align*}
p_\ast^{(CTQW,4)}(\pi) >\ &p_\ast^{(DTQW,4)}(\pi) > p_\ast^{(RW,4)}(\pi), \\
p_\ast^{(CTQW,4)}(2\pi) <\ &p_\ast^{(DTQW,4)}(2\pi) < p_\ast^{(RW,4)}(2\pi).
\end{align*}
The greater $p_\ast^{(W,4)}(\pi)$ is, the better the model reflects the given function $y=f(x)$. The smaller $p_\ast^{(W,4)}(2\pi)$ is, the better the model reflects the given function $y=f(x)$. Thus, we conclude that CTQW is the best walk for extrapolation in three walks.
We should note that the above all values,
\begin{align*}
\left\{ p_\ast^{(W,4)}(\pi),\ p_\ast^{(W,4)}(2\pi) : W \in \left\{ CTQW,\ DTQW,\ RW\right\}\right\},
\end{align*}
are different.
%%%%%%%%%%%%%%%%%%%%%%%%%%%%%%
%\section*{Acknowledgments}

%\acknowledgement
%We would like to thank Masato Takei and Kei Saito for valuable comments.

%%%%%%%%%%%%%%%%%%%%%%%%%%%%%%

%-------------------------------------------------------
% @.@- Write the author's addresses as below.
% @.@- Use \\ to split two or more authors.

%\color{black}
%\begin{address}
% Norio Konno \\
% Department of Applied Mathematics, Faculty of Engineering \\ 
% Yokohama National University \\
% Hodogaya, Yokohama, 240-8501, JAPAN \\
% E-mail address: \texttt{konno-norio-bt@ynu.ac.jp} \\
%\\
%Shohei Koyama \\
%Graduate School of Science and Engineering \\ 
% Yokohama National University \\
% Hodogaya, Yokohama, 240-8501, JAPAN \\
%E-mail address: \texttt{koyama-shohei-mw@ynu.jp}
%\end{address}

\end{document}